\documentclass[12pt]{article}
\begin{document}
\pagestyle{plain} \setcounter{page}{1}
\begin{center}
{\large \textbf{Comment on the Variation of Fundamental Constants}}
\vskip 0.2 true in
{\large J. W. Moffat}\footnote{e-mail: john.moffat@utoronto.ca}
\vskip 0.2 true in
\textit{Department of Physics, University of Toronto, Toronto, Ontario M5S
1A7, Canada} \vskip 0.2 true in and
\vskip 0.2 true in
\textit{Perimeter
Institute for Theoretical Physics, Waterloo, Ontario N2J 2W9, Canada}
\vskip 0.2 true in
\begin{abstract}
It is argued by Duff that only the time variation of dimensionless
constants of nature is a legitimate subject of enquiry, and that
dimensional constants such as $c,\hbar, G...$ are merely human constructs
whose value has no operational meaning. We refute this claim and point out
that such varying dimensional ``constants'' can have significant physical
consequences for the universe that can be directly measured in experiments.
Postulating that dimensional constants vary in time can significantly
change the laws of physics. \end{abstract}

\vskip 0.3 true in  \end{center}

\date{\today}

In a recent article, Duff~\cite{Duff,Duff2} has asserted that dimensional
constants such as $c,\hbar, G...$ are human constructs that have no
intrinsic meaning in physics. He states that the dimensional constants
``are merely human constructs whose number and values differ from one
choice of units to the next and which have no intrinsic physical
significance.'' Of course, as long as these dimensional constants {\rm
remain constants}, then we can set them equal to unity, and treat them as
a means to change units. However, once we postulate that these
constants are no longer really ``constant'' but vary in space
and time, we can no longer assert that they are just ``human
constructs'' that allow us to change from one set of units to
another. It is this source of confusion that I wish to address
in the following.

If we consider the fine structure
constant, $\alpha=e^2/\hbar c$, then clearly varying the electric charge
$e$ with time will have quite significant consequences compared to varying
$\hbar$ or $c$. Dirac was one of the first
physicists to suggest that, in connection with his theory of
large numbers, fundamental dimensional constants may vary in time during
the expansion of the universe~\cite{Dirac}. Indeed, he considered that
Newton's gravitational constant $G$ varied with time. If one so wishes,
one can consider the measurable quantity, ${\dot G}/G$, in which the only
dimensional quantity that enters the formula is the time $t$ and $t$ is
measured by standard clocks.

The idea that a variation in the speed of light $c$ would have
significant consequences for cosmology was first suggested in the context
of the big bang model a decade ago~\cite{Moffat}. It was followed by other
suggested cosmological models~\cite{Magueijo,Clayton} and
has been the subject of attention by physicists in investigations of extra
dimensions, strings and branes~\cite{Youm}. The media has recently been
drawn to the importance of these ideas by the possible variation of
$\alpha$ to one part in 100,000, deduced from observations of QSO
absorption lines~\cite{Webb}. The significance of these observations and
the consequences of a variation of $\alpha$ has been the subject of
intensive research~\cite{Dvali,Moffat2}. To these investigations can be
added the recent paper by Davies et al.,~\cite{Davies}. It seems clear
that whether you vary $e$, $\hbar$ or $c$ will have very different
consequences for physics. Such consequences can be detected and measured
and from these results, we can decide which ``dimensional'' constant of
the three involved is varying, even though the effects of a varying
$\alpha$ appear to be falsely ``hidden'' in the variation of either $e$,
$\hbar$ or $c$. Considering the variation of $\alpha$ in isolation from
the rest of physics and not taking into account the variation of either
$e$, $\hbar$ or $c$ individually seems an unacceptable approach to the
problem.

In a paper by the author~\cite{Moffat2}, it was
shown that varying $e$ would significantly violate Einstein's weak
equivalence principle, unless some very exotic features of dark matter were
invoked\footnote{see also, e.g.~\cite{Dvali}}. It was concluded by the
author that it only seemed reasonable to vary $c$ and that $c$ should
increase as you went into the past in the expanding universe, i.e. varying
$c$ by the amount suggested by the observations of Webb et al. would not
obviously violate any current experiment, although it would require a major
revision of relativity theory. It was pointed out that such a
variation in $c$ would become critical at big bang nucleosynthesis at a
red shift $z\sim 10^9-10^{10}$. Future analysis of satellite observations
could significantly restrict a variation in $\alpha$~\cite{Martins}.
Bekenstein,~\cite{Bekenstein} in his early paper on a model of the
variation of the electric charge $e$, came to the same conclusion that
varying $e$ would produce a strong violation of the weak equivalence
principle.

It is clear (also to the media) that the consequences of a varying $c$
could significantly alter our description of the physical world, for the
second postulate of Einstein's special relativity: ``The speed of light is
the same constant for all observers irrespective of their motion and the
motion of the source'' would no longer hold true. Consider
$c(t)=c_0\phi(t)$, where $c_0$ is the standard measured speed of light
$c_0=299792458\,m\,s^{-1}$, then writing $c(t)/c_0=\phi(t)$ shows that the
dynamical effects of the dimensionless $\phi(t)$ in a postulated action for
a theory describing the universe will obviously change our predictions
from those in which $c(t)=c_0$ and $\phi(t)=1$. {\it We have to
fundamentally alter our understanding of spacetime and gravitation}. This
could be done by {\it breaking the symmetry of Lorentz invariance} in the
action,~\cite{Moffat,Magueijo} or e.g. postulating the existence
of two spacetime metrics connected by the gradient of a scalar
field, which would introduce two varying light
cones~\cite{Clayton}. Once we postulate that the speed of light
varies in time, then we must somehow change Einstein's special
theory of relativity in a fundamental way.

Varying other dimensional constants will be expected to have similar
significant consequences for the laws of physics and produce theoretical
models in which the different assumed actions predict differing physical
results. If we assume that $\hbar$ varies in time with $c$ kept
constant, this would produce detectable effects in atomic spectra but
it would not obviously alter quantum mechanics at a fundamental level, nor
would it require a revision of special relativity.

In conclusion, we realize that postulating that constants such
as $\hbar$, $c$ or $G$ vary in time elevates them to a status of
fundamental importance in theoretical physics, and makes them more than
merely human constructs whose number and values differ from one choice of
units to the next with no operational meaning.

\vskip0.2 true in \textbf{Acknowledgments} \vskip0.2 true in This
work was supported by the Natural Sciences and Engineering Research
Council of Canada. \vskip0.5 true in

\end{document}